# Tunable macroscopic defect patterns induced by a low-frequency AC electric field in ferroelectric nematic liquid crystals

Natalia Podoliak*, Lubor Lejček, Martin Cigl, and Vladimíra Novotná

*Institute of Physics of the Czech Academy of Sciences, Na Slovance 1999/2, 182 00 Prague 8, Czech Republic*

Corresponding author*: Natalia Podoliak, e-mail: podoliak@fzu.cz

**Abstract**

Regulation of topological structures and pattern formation is attracting wide interest in the field of condensed matter. Liquid crystals (LCs) represent soft matter with a remarkable combination of fluidity and anisotropic properties. Topological defects may appear in confined LCs under external stimuli. Recently discovered ferroelectric nematics (NF) opened exceptional opportunities in technologies owing to high permittivity and polarisation. Polar properties of NF supply more variability to topological structures. In this research, we present tunable 2D topological defect arrays in NF compound, induced by an alternating (AC) electric field in simple sandwich cells without pre-patterning. The observed arrays of defects form pseudo-square lattices, which character and periodicity depend on the frequency of the applied field and partially on the cell thickness. The observed effect is explained to occur due to the competition between elastic and electrical forces. The proposed system can be useful to create reconfigurable spatially periodic polarisation structures.

**Introduction**

Regulation of topological structures and pattern formation is attracting wide interest in the field of condensed matter and materials science[1]. Whereas in solid state there are promising applications of topological insulators in spintronic devices[2], quantum computers,[3] advanced magnetoelectronic and optical devices[4], topological soft matter opens new perspectives towards modern technologies due to interconnections between self-assembly and geometry[5,6]. Moreover, owing to larger-scale effects in soft systems, their study can serve as a tool for better understanding of the effects in other topological systems.

Liquid crystals (LCs) as self-assembly of organic molecules are known for their remarkable combination of fluidity and anisotropic properties. As a result of intermolecular interactions, thermotropic LCs can be organised in various phases within specific temperature ranges and sensitively reflect any external fields and conditions. Behaviour of LC systems primarily depends on the confinement conditions and various kinds of defects can occur. For liquid crystal research, sandwich cells are utilised for electro-optical measurements. Such cells are constructed of two parallel glass plates with transparent conductive electrodes to apply electric field across the cell. The orientation of the molecules and polarisation in the cell is affected by the molecular alignment at the surfaces. Two basic configurations can be realised via a surfactant layer at the glass surfaces: planar alignment with parallel preferred molecular orientation (rubbing direction) ensures homogeneous geometry (HG); in homeotropic geometry (HT), the molecules are arranged perpendicular to the surfaces.

Recently, ferroelectric nematic phase ($N_F$) was experimentally realized[7-9], a century after its prediction[10,11]. In classical nematic phase (N), which can appear below the isotropic liquid (Iso) phase on cooling, there is a long-range orientational ordering with the prevailing molecular orientation described by a unit vector named director, *n*, having an inversion symmetry. In the $N_F$ phase, the director represents an axial polar vector, and the head-to-tail symmetry is broken. Consequently, the resulting phase is polar with spontaneous polarisation being locally parallel to *n*. Since the $N_F$ phase discovery, ferroelectric nematogens have been intensively studied[12-22]. For these materials, very large values of polarisation as well as electric susceptibility have been confirmed[12-18,23,24]. Additionally, non-linear properties, namely efficient second harmonic generation (SHG) supporting polar character of the $N_F$ phase, and electro-optic response with exceptionally low threshold values have been reported[16,25-27], giving these materials huge applicational potential.

Generally, all LC systems are prone to defect formation within their structure. In homogeneous state, defects are not stable and tend to annihilate and minimise the free energy of the system. However, in confined geometry, topological defects arise naturally due to discontinuity of ordering at the surfaces. Disclination lines are present in nematics if the director makes continuous rotation along a closed loop. Previously, formation of 2D defect structures in a sample with one free surface under the influence of magnetic field was described by de Gennes[28]. As defects can be observed in specific geometries with different upper and lower surfaces, a regular network of singular points was observed at the nematic-isotropic interface[29], resulting in oblique director orientation. Additionally, defect arrays can be induced by external stimuli. For example, an appearance of a chiral stripe pattern was induced in an achiral material by subjecting it to a pressure-driven flow in a microfluidic cell[30]. The possibility to influence the molecular orientation by electric field and pattern formation is highly advantageous and challenging[31,32]. The appearance of 2D periodic patterns under the influence of AC electric field was mentioned for a nematic in planar geometry[33]. Large-scale topological defect networks were induced by AC electric field with an optical tweezer technique in a nematic doped with ionic impurities[34]. Additionally, arrays of micropillars modifying the cell surface were utilised for a pattern stabilisation[35].

The mentioned above topological structures appear due to a system frustration and under very specific conditions, with limited control of parameters. A different mechanism is related to orientationally pre-patterned surfaces[36], which depends on the advanced technology of photopatterning. However, the parameters of the pre-patterned configurations cannot be tuned. Recently, periodic polarization structure in a photopatterned sample was reported for the ferroelectric nematic LC giving a deeper insight to the flexoelectric coupling in the director field for such a type of materials[37].

In this paper, we demonstrate the possibility to create tunable 2D reconfigurable topological structures in ferroelectric nematics confined in simple sandwich cells without any pre-patterning, by the application of an AC electric field. The induced defects form pseudo-square lattices, which partially persist after the applied field is switched off. The periodicity of the structure can be changed by varying the field characteristics (frequency). We investigate the behaviour of the defect structures in dependence on the cell parameters and the applied electric field characteristics. We propose a theoretical explanation based on the competition between the elastic and electrical forces. Our findings could be useful for the design of patterns with reconfigurable spatially periodic polarisation structures.

**The patterns generation and observation**

For the current investigation, a ferroelectric nematic material designated NF6 is utilised, exhibiting a direct transition from Iso to $N_F$ phase at 65 °C. The synthesis and mesomorphic characteristics of the material were published previously[15]. The studied material was filled into commercial glass cells with transparent indium tin oxide (ITO) electrodes and surfactant layers ensuring the necessary geometry (HT). The low-frequency electric field (1-500 Hz) of sinusoidal or triangular profile was applied to the sample perpendicular to the surfaces and, hence, along the molecular long axis in HT geometry.

The virgin texture observed under a polarised light optical microscope (POM) for a 5-μm HT sample in $N_F$ phase after cooling from Iso represented a random distribution of coloured domains of irregular shapes missing any periodical features, see Supplementary Fig. S1. After the application of an AC electric field, a periodic array of defects was induced. The pattern covered the whole electrode area, with inhomogeneities close to the electrode edges (Fig. 1a). After switching off the field, the pattern persisted for some time, depending on the temperature at which the sample was hold and, hence, the material viscosity. At room temperature (RT), the pattern remained till the next day. However, it became smashed and contained additional smaller domains within the initial pattern (Supplementary Fig. S2). For higher temperature and lower material viscosity, the pattern degeneration was faster. In Fig. 1b-d one can see the evolution of the pattern with time for T=40°C, with the gradual appearance of the additional defects within the initial structure and the pattern degradation as a result.

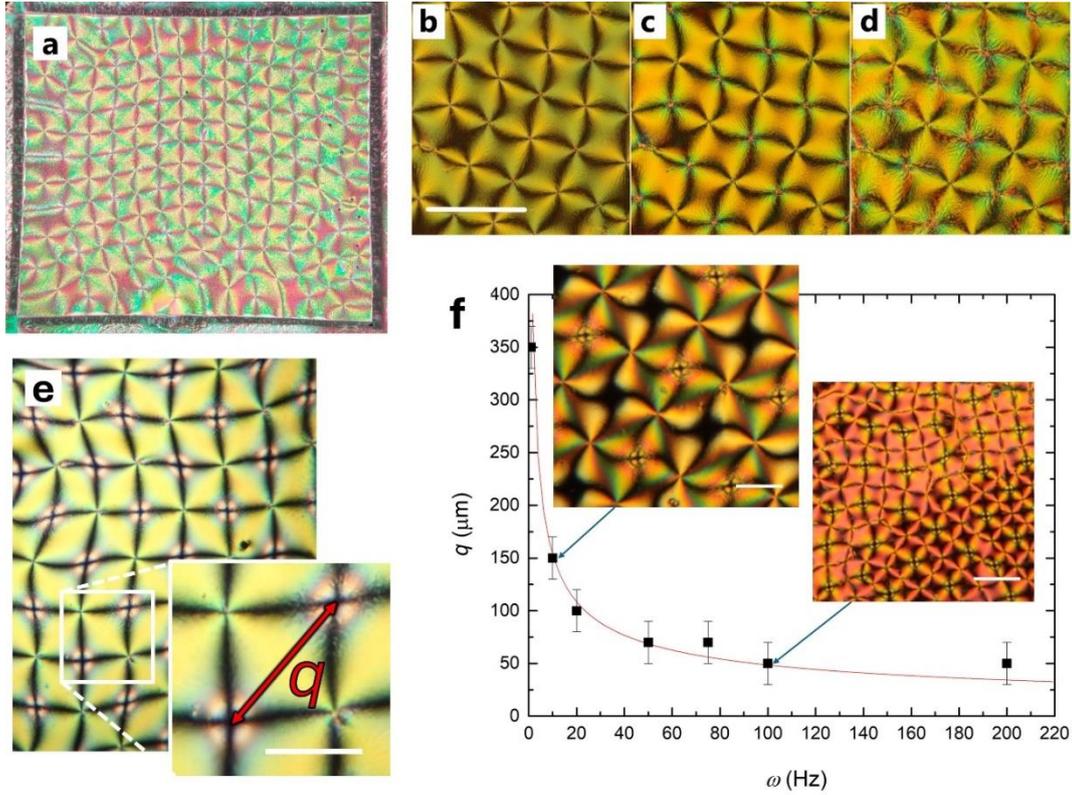

**Fig. 1| Square-lattice patterns observed for a 5-μm HT sample in the $N_F$ phase under the influence of an AC electric field. a** POM image, taken at RT, of the 2D square-lattice system of defects covering the whole electrode area with observed inhomogeneities close to the electrode edges, after the application of 1 Hz AC electric field. The size of the electrode area is 5mm×5mm. Crossed polariser and analyser are oriented along the edges of the photo. **b-d** Microphotographs taken at T=40°C **b** under the applied AC electric field of 100 Hz, **c** immediately after and **d** 20 seconds after the field was switched off. **e** The pattern obtained using 10 Hz AC field. Two types of defects are observed, switchable (upper right and bottom left in the inset) and unswitchable (upper left and bottom right in the inset). Mean distance between the defects of similar type, $q$, is shown in the inset. **f** The dependence of $q$ on the frequency of the applied electric field, $\omega$. The points are experimentally measured values while the curve corresponds to the fitting of Eq. (9). Additionally, the pattern textures obtained for certain frequency values are presented. Crossed polariser and analyser are along the edges of the photos. Scale bars correspond to 100 μm.

We analysed the textures for various frequencies from 1 Hz to 500 Hz and established that the periodicity of the patterns reveals strong dependence on the frequency of the applied AC field. Indeed, for 1 Hz the macroscopic pattern can be observed even by naked eye, as it is presented in Supplementary Fig. S3. With the frequency change, a defect movement and reconfiguration took place with the consequent formation of a new pattern. For higher frequencies, the number of the defects increased and the distance between them decreased. And vice versa, with the change from higher frequencies to lower, the distance between the defects increases. In Supplemental information, the videos presenting the pattern structure changes with frequency are presented. For higher frequencies, the pattern reconfiguration was slower, which might relate to a larger number of the defects and, hence, higher deformation energy of the system. The periodicity of the patterns induced by 100 Hz and 200 Hz applied fields practically did not differ. For frequencies higher than 200 Hz, the textures became massed and missing any periodic features distinguishable by POM. The pattern periodicity did not depend on temperature, at which it was induced, but at lower temperatures the molecular reorientation was slower, and the periodic structure was generated after a longer time, reaching tens of seconds at RT. We have not found relevance between the amplitude of the applied field and the periodicity of the defects. The threshold field of the effect has been established to be about 1 V/μm.

We discovered that under the AC field application we can induce a pseudo 2D square-lattice system of defects. By analogy with ordinary nematics[38], we can describe such a type of defects structure as a system of ($\pm 2\pi$)-wedge disclinations. During the microscopic investigation of the patterns under the applied field, we observed that the defects are not all equal and two types of defects are present, differing in their response to the applied field. The difference is clearly visible for slower frequencies, and it consists in the fact that one type of the defects is changing under the applied AC field, while the other type remains unchanged (see, for example, Supplementary Video 2). We will call the first type switchable defects, and the second one unswitchable defects. In Fig. 1e, the pattern created using 10 Hz applied field is presented. At this frequency, the difference between the mentioned above defects is visible. In the inset of Fig. 1e, an enlarged fragment of the pattern showing one unit cell of the structure is highlighted. In the upper right and bottom left corners of the square unit cell the switchable defects are situated; subsequently, unswitchable defects are observed in the upper left and bottom right corners. The reason for the different defect behaviour and the nature of the defects will be speculated later in the Defect description and theoretical explanation section. It should be also mentioned that, as it is seen from Supplementary Video 2, the neighbouring switchable defects are switching out of phase, which will also be explained in the Defect description and theoretical explanation section.

We marked the distance between two neighbouring defects of the same type (switchable or unswitchable) as $q$ (see the inset of Fig. 1e) with the aim to quantitatively describe the observed dependence of the pattern periodicity on the applied field frequency, $\omega$. We established $q$ for the patterns obtained at different values of $\omega$ and the results of the measurements are presented in Fig. 1f (points with error bars). Based on the domain orientation under the polarised light microscope, the director profile of the pattern structure can be deduced. The domains orientation indicated the radial director distribution near the defects. To prove it, we acquired images with the inserted at 45° full-wave retardation plate ($\lambda$-plate) and quarter-wave retardation plate ($\lambda/4$-plate, changing the light polarization state into circular), $\lambda$=530 nm (Fig. 2). The retardation plates allowed us to distinguish between the regions with different molecular orientation by the colour change.

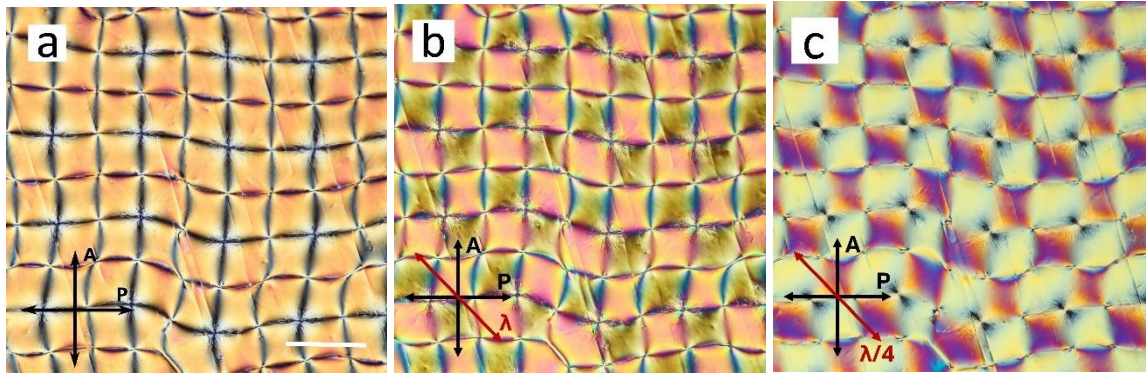

**Fig. 2| POM images of the pattern structure generated by 50 Hz AC electric field in the studied sample. a** The image obtained without additional optical systems. **b** The image obtained with an inserted at 45° full-wave retardation plate (λ-plate). **c** The image obtained with an inserted at 45° quarter-wave retardation plate. The polariser and analyser are oriented along the edges of the photos and are shown with black arrows; the retardation plates are shown with red arrows. The scale bar corresponds to 100 µm.

Additionally, we studied the effect of the cell thickness on the defect array formation. Except for the previously studied 5-µm cell, we examined various thicknesses: 1.5, 3, 7 and 12 µm. For 3-µm and 7-µm cells, the textures behaviour resembled those described above for the 5-µm cell, with analogous parameters. However, for the 3-µm cell, the pattern formation process was slower and regular patterns appeared after longer periods of time than for the thicker cells, at least for tens of seconds. For thin 1.5-µm cell, the pattern formation was strongly clamped by the cell surfaces influence and the array of defects was not uniform.

Further, we examined the effect of AC field in another cell geometry, namely for planar HG cells with antiparallel rubbing at the surfaces. The virgin textures contained twisted domains (Supplementary Fig. S4), as it was described in the previous research[39]. In this geometry, the applied electric field should overcome the surface anchoring and reorient the molecules at the surfaces. As in the $N_F$ phase due to very high responsiveness to an
9

electric field the molecules could be easily reoriented[26], the virgin HG geometry was transformed into the HT one by the applied voltage. Under an AC field, for the cells of the thickness 3 µm and higher, we succeeded in creating similar defect patterns as in the HT cells. For thin 1.5-µm HG cell, the threshold field for the 2D square pattern creation increased up to 2.5 V/µm, and the pattern was not regular (Supplementary Fig. S5). For the smaller field of 1-2.5 V/µm, stripe textures were observed perpendicular to the rubbing direction (Supplementary Fig. S6) for frequencies 10 Hz and higher. At a small frequency of 1 Hz, any periodic features were not observed even for the field higher than 2.5 V/µm. Apparently, for thin 1.5-µm HG cell, the influence of the surfaces dominated.

**Defect description and theoretical explanation**

To elucidate the effect of the applied electric field in the ferroelectric $N_F$ phase, let us point out the differences between the $N_F$ phase and ordinary nematics relevant to the molecular organization at the sample surfaces. Due to the polarity of the molecules, there are different interactions of the molecular heads and tails with the surfactant layers[40]. While in ordinary nematics the director configurations in aligned cells can be uniform, in the $N_F$ phase an additional twist (in planar cells with antiparallel alignment) or splay (cells with homeotropic boundary conditions) will appear due to polar surface anchoring.

We suppose that the electric energy applied to the sample is absorbed in the system resulting in the creation of an array of defects. Additionally, defects can vibrate due to the interaction of the applied electric field with the spontaneous molecular polarisation and the molecular reorientation brings about bound charges within the sample bulk. In general, the system of defects is unstable, but it is stabilised by an applied electric field.

In Fig. 3a, we present the texture of the sample in the 5-µm HT cell under the applied field of the frequency 10 Hz, augmented with the schematic expected director orientation shown by arrows, and, consequently, the molecular dipole moment arrangement. We consider that switchable defects are (+2π) - disclinations, with either positive or negative charge (red or blue points in Fig. 3a); unswitchable defects correspond to (-2π) – disclinations, they are not charged and are shown with white points.

The switching is related to the director reorganisation around the defects, and charged defects compensate their charges at longer distances with their oppositely charged neighbours. Based on our observations, we can propose a schematic molecular orientation within the cell thickness for the charged defects, shown in Fig. 3b. Coordinate axes are chosen with (x,y)-plane to be parallel to the surfaces and z-axis across the sample. Parameter $q$ is the mean distance between the neighbouring oppositely charged defects. The core regions of the charged defects are along (±z) – directions with singular points (red and blue points) within the defects. The charged defect length is marked as $l_z$ (dashed lines in Fig. 3b). We suppose that the charged defects vibrate along ($\pm z$) direction while the vibration of the uncharged defects is negligible. The dynamics of the molecular reorganisation around the defects in response to the applied AC electric field is visualised in Supplementary Video 4 and Video 5. The video animations schematically demonstrate the vibrations of the charged defects (Video 4) and only slight reorientation of the molecules near the uncharged ones (Video 5).

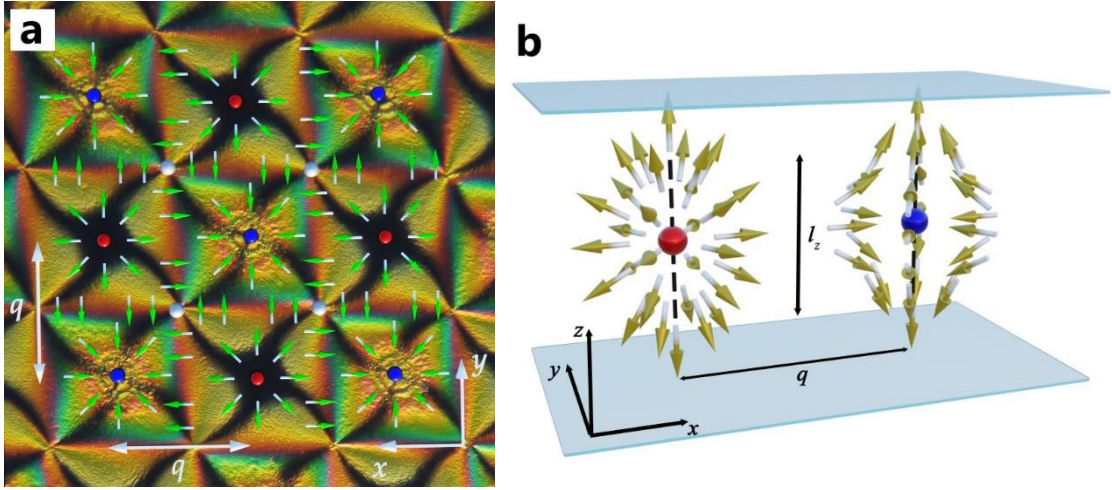

**Fig. 3| Defect structure. a** The texture of the sample in the 5-μm HT cell under the applied field of the frequency 10 Hz and the schematic orientation of the molecules; arrows denote the molecular director and, consequently, the molecular dipole moment orientation. Coloured points are $(+2\pi)$- disclinations while white points are $(-2\pi)$- disclinations. Evidently, $(-2\pi)$- disclinations are uncharged while $(+2\pi)$- disclinations have either positive (red point), or negative (blue points) charges. **b** Schematic molecular orientation near the charged defects in the sample with HT polar anchoring geometry. Parameter $q$ is the mean distance between the neighbouring oppositely charged defects and the charged defect length is marked as $l_z$.

When an external AC electric field with the frequency $\omega$ and amplitude $E_o$ is applied along the normal to the sample surfaces, it interacts with the spontaneous polarisation $\overrightarrow{P_S}$. The electric energy can be absorbed in the sample, which stimulates the creation of the defect system and its stabilisation. The charged defects vibrate between the upper and lower glass plates along the z-axis (Fig. 3b) with the time dependent velocity $v = \dot{z}$. If we consider $\Phi$ to be the dissipated energy per unit time and take the defect density $\frac{1}{q^2}$, $\Phi$ can be written as $\Phi = -F_S v$ with the friction force $F_S = -\pi l_z \gamma_1 \ln\frac{R}{r_c} v$, which was calculated for the nematic phase in the literature[41]. Here $R$ is the outer radius of the surface surrounding the disclination, $r_c$ is the core radius, and parameter $\gamma_1$ is the nematic rotational viscosity. The force corresponding to the electric field interaction with the liquid crystal polarisation can be written in form $-(\pi R^2)(P_S E_o \cos\alpha_s)\cos(\omega t)$, where $(\pi R^2)$ is the disclination area in $(x,y)$-plane. We take $\alpha_s$ as a geometrical factor to describe the mean molecular inclination over the defect length $l_z$. We can expect that in general $l_z$ depends on the frequency of the applied field. Nevertheless, for the simplicity of calculations, we consider it to be constant in the model. The energy of the electric field applied to the sample leads to the creation and stabilisation of the defect system due to the dissipation of the electric energy. At a stationary state, the energy balance between the field energy dissipated in the system per period $\tau = \frac{2\pi}{\omega}$ plus the disclination self-energy, $E_S$, multiplied by the defect density, will be in equilibrium with the energy of the electric field:

$$\frac{1}{q^2}\left[\int_0^\tau \Phi dt + E_S\right] = \frac{1}{\tau}\int_0^\tau |hP_S E_o \cos\alpha_s \cos(\omega t)|\, dt \qquad (1)$$

The total disclination self-energy of the charged disclinations, $E_S$, consists of the elastic energy, $E_{es}$, and the electric contribution, $E_Q$:

$$E_S = E_{es} + E_Q \qquad (2)$$

The elastic energy of the disclinations can be written in a form described in the literature for nematics[28,41]:

$$E_{es} = l_z \pi K \ln\frac{R}{r_c}, \qquad (3)$$

where $K$ is the elastic constant in one-constant approximation. To estimate the electric interactions of the charged defects in the sample, the disclination self-energy due to the disclination charge can be expressed in analogous way to the one described in the literature for classical electrodynamics[42]:

$$E_Q = \frac{1}{2\varepsilon_o\varepsilon_r} \iint \frac{\rho(\vec{r})\rho(\vec{r'})}{|\vec{r}-\vec{r'}|} dV dV' \approx \frac{Q^2}{2\varepsilon_o\varepsilon_r l_z^2} \int_{-\frac{l_z}{2}}^{\frac{l_z}{2}} dz \int_{-\frac{l_z}{2}}^{\frac{l_z}{2}} dz' \frac{1}{\sqrt{r_c^2+(z-z')^2}} \quad (4)$$

where $\varepsilon_o$ and $\varepsilon_r$ are the dielectric permittivity of vacuum and the relative permittivity, respectively. The volume density of charges $\rho(\vec{r})$ is $Q/V$, where the volume $V$ contains one disclination, i.e. $V \sim \pi R^2 l_z$. Configurational disclination charge $Q$ can be determined as $Q \approx \pi(R-r_c)l_z P_S \sin\alpha_s$, as it was described for the disclinations in ferroelectric smectics[43]. As $l_z > r_c$, the above double integral can be evaluated as $\sim l_z \ln\frac{l_z}{r_c}$ and the self-energy of the charged disclination of the length $l_z$ can be estimated as $E_Q = \frac{Q^2}{2\varepsilon_o\varepsilon_r l_z} \ln\frac{l_z}{r_c}$. We take as the approximated total disclination self-energy the expression:

$$E_S = l_z \pi K \ln\frac{R}{r_c} + \frac{Q^2}{2\varepsilon_o\varepsilon_r l_z} \ln\frac{l_z}{r_c} \quad (5)$$

To obtain the dynamic equation, we may introduce the temporary force $-\pi l_z \gamma_1 \ln\frac{R}{r_c} \dot{z}$ and the dynamical equation takes a form:

$$-l_z \pi \gamma_1 \ln\frac{R}{r_c} \dot{z} - (\pi R^2)(P_S E_o \cos\alpha_s)\cos(\omega t) = 0 \quad (6)$$

We expect the charged defect follows the electric field frequency and equation (6) gives the velocity $\dot{z}$ of the disclination vibration. The expression $\int_0^\tau \Phi \, dt$ can be then evaluated as

$$\int_0^\tau \Phi \, dt = \pi l_z \gamma_1 \ln\frac{R}{r_c} \int_0^\tau (\dot{z})^2 dt = \frac{\pi}{\omega} \frac{(\pi R^2)^2 (P_S E_o \cos\alpha_s)^2}{\left(\pi l_z \gamma_1 \ln\frac{R}{r_c}\right)} \quad (7)$$

One can find the detailed calculations in Supplemental information. From the calculations, the obtained distance between the defects is:

$$q(\omega) = \sqrt{\frac{\pi}{2}} \sqrt{\frac{(\pi R^2)^2 (P_S E_o \cos\alpha_s)}{\omega h l_z \gamma_1 \ln\frac{R}{r_c}} + \frac{E_S}{h(P_S E_o \cos\alpha_s)}} \qquad (8)$$

We simplified such a dependence on the applied field frequency to the expression:

$$q = \sqrt{\frac{a}{\omega} + b} \qquad (9)$$

and fitted to this equation the experimental data demonstrated in Fig.1f (red curve in Fig. 1f). This frequency characteristic reflects the fact that at higher frequencies the defects cannot follow the oscillations of the applied field.

**Discussions and conclusions**

In the present research, we have induced the arrays of defects in the ferroelectric nematic liquid crystal in simple sandwich cells without any pre-patterning on the cell surfaces, by the application of a low-frequency oscillatory electric field. We have established the conditions and the defect pattern parameters. We suppose that the electric energy applied to the sample is absorbed resulting in the system of charged and non-charged defects. The observed defects are ($\pm 2\pi$)-disclinations. In response to the AC electric field, the charged defects oscillate along z-axis. The uncharged defects are of elastic origin only, whereas the self-energy of the charged defects additionally includes the charge energy and the vibrational energy. We can neglect the interaction energies between the defects as the distances between them are of the order of hundreds of micrometres.

We expect that the studied defect structures have zero total polarisation as the local molecular dipole moments are mutually compensated. We have proposed a model to describe the dynamic equilibrium and derived the dependence of the defect structure parameter, the distance between the defects, $q$, on the applied AC field frequency. We have fitted the experimental values and found that, despite the model simplifications, we have reached a good consistency between the proposed frequency dependence of the structure parameters and the experimental data (Fig. 1f). The presented explanation of the observed effect represents an original approach to solve similar effects in various self-assembling systems and can be applied in a generalised form in other areas of soft matter as well as biological systems.

The term active matter is relevant to a system out of thermodynamic equilibrium as a result of the energy input[44]. By the transformation of the applied electric field energy into the quasi-equilibrium defect array structure, the studied system might be considered as active. In active nematics[45], topological defects are generated and annihilated, bringing about new functionalities. Most often, the observed topological defects appear chaotically. In our case, the elastic properties of the material and its responsiveness to the applied electric field allows for the controllable and tuneable defect structure. This might offer opportunities for the development of entirely new technologies in such branches as, for example, soft robotics.

**Methods**

For our research, we utilised ferroelectric nematic liquid crystalline compound NF6, synthesis and mesomorphic properties of which were published previously[15]. We used sandwich-type commercial glass cells, purchased from WAT PPW company, Poland, with transparent indium tin oxide (ITO) electrodes and surfactant layers ensuring homeotropic (HT) or homogeneous (HG) geometries. Various cell thicknesses: 1.5, 3, 5, 7 and 12 µm were used to investigate the dependence of the obtained structures on the cell gap. The electrode area of the cells was 25 mm$^2$ for all the samples. Besides, a cell with a larger electrode area (64 mm$^2$) was utilised for comparison.

The cells were filled with the material in the isotropic (Iso) phase by means of capillary action. The textures were investigated under the polarising light microscope Nikon Eclipse E600, equipped with a Linkam heating-cooling stage with temperature stabilisation ±0.1 K, and a temperature controller Linkam TMS 94. A photo camera Canon EOS 700D was used for textures photo and video acquisition.

**Data availability**

Data relevant to this study could be found in Supplementary information. Additional information could be provided by corresponding author upon request.

**Acknowledgment**

Authors acknowledge the project 24-10247K (the Czech Science Foundation).

**Authors contributions**

V.N. supervised the project; N.P. conducted the experimental work and data analysis; L.L. developed the theoretical model and carried out the calculations; M.C. is responsible for the materials studied; N.P., V.N. and L.L. prepared the manuscript.

**Competing interest**

The authors declare that they have no competing interests.

**Additional information**

The online version contains supplemental material available at